\documentclass[showpacs,preprintnumbers,amsmath,amssymb,12pt,floatfix,epsfig]{revtex4}
\usepackage[usenames, dvipsnames]{color}
\usepackage{amssymb}
\usepackage{bm}
\usepackage{graphicx}
\usepackage{revsymb}
\usepackage{amsmath}
\usepackage{dcolumn}
\newcommand{\be}{\begin{eqnarray}}
\newcommand{\ee}{\end{eqnarray}}

\begin{document}
\draft
\title{Competition of deformation and neutron-proton pairing in Gamow-Teller transitions for $^{56, 58}$Ni and $^{62,64}$Ni}

\author{Eunja Ha \footnote{ejha@ssu.ac.kr} and Myung-Ki Cheoun \footnote{Corresponding Author: cheoun@ssu.ac.kr}}
\address{Department of Physics and Origin of Matter and Evolution of Galaxy (OMEG) Institute, Soongsil University, Seoul 156-743, Korea}

\begin{abstract}
We investigated effects of neutron-proton ($np$) pairing
correlations comprising isoscalar (IS) and isovector (IV) component on Gamow-Teller (GT) transitions of $^{56, 58,62,64}$Ni by taking into account
nuclear deformation in a deformed Woods-Saxon mean field. Our GT calculations based on a deformed Bardeen-Cooper-Schriffer (DBCS) approach have been performed within a deformed quasi-particle random phase approximation (DQRPA) which
explicitly includes the deformation as well as all kinds of pairing correlations not only at the DBCS but also at the DQRPA stage. Competition of these two effects, the deformation and the IS spin-triplet and IV spin-singlet component of the $np$ pairing, is shown to be significant for understanding the GT strength distributions of Ni isotopes.
\end{abstract}

\date{\today}

\maketitle

\section{Introduction}
In general, like-pairing correlations, such as proton-proton ($pp$)
and neutron-neutron ($nn$) pairing usually adopted in understanding the nuclear superfluidity, have isovector spin-singlet ($T=1, J=0$) mode and manifest themselves as nuclear odd-even mass staggering. They contribute to keeping spherical properties due to their $J =$ even couplings against the deformation treated in the mean field.

On the other hand, neutron-proton ($np$) pairing correlations in the residual interaction are expected to play meaningful roles in $N \simeq Z$ nuclear structure, and relevant nuclear electro-magnetic (EM) and weak transitions because protons and neutrons in these nuclei occupy the same (or nearby) orbitals leading to the maximum spatial overlap.

The $np$ pairing correlations have two different modes, viz. isoscalar (IS) spin-triplet ($T=0, J=1$) and isovector (IV) spin-singlet ($T=1, J=0$) \cite {Chen-Gos, Wolter, Goodman,simkovic03}. The IV $np$ pairing part, whose spin-singlet gives rise to anti-aligned pair, can be investigated by the study of $T=0$ and $T=1$ excited states in even-even and odd-odd nuclei. But the IS $np$ pairing part has a spin-aligned pair structure. For example, deuteron ground state as well as $np$-scattering phase shift analyses indicate strong attractive $np$ pairing features due to the strong $^{3}S_{1}$ tensor force. Inside nuclei, they are believed to contribute mainly to the mean field. But, even after converted to a mean field, it is quite natural to conjecture that there should still remain such strong attractive interactions due to the $np$ pairing in the residual interaction, although direct data implying the IS $np$ pairing inside nuclei are still controversial even in the $N \simeq Z$ nuclear structure study

Recently, with the advent of modern radioactive beam facilities producing the $N \simeq Z$ nuclei, competition or coexistence of these two IV and IS $np$ pairing in the residual interaction for $N \simeq Z$ nuclei is emerging as an interesting topic in the nuclear structure. Detailed reports about the recent progress regarding the $np$ pairing correlations in nuclear structure can be found at Refs. \cite{Sweden16,Brend16}. In particular, the IS spin-triplet part in the $np$ pairing has been argued to be very elusive compared to the IV spin-singlet part stemming from the like- and unlike-pairing correlations. Moreover, the deformation may also affect the IS and IV pairing correlations or vice versa because the IS $np$ pairing has $J=$ odd coupling inducing (non-spherical) deformation contrary to the IV $np$ mode keeping $J=$ even coupling.

Importance of the $np$ pairing has been discussed two decades ago in our early reports of the single- and double-$\beta$ decay transitions \cite {Ch93,pan} within a spherical quasi-particle random phase approximation (QRPA) framework with a realistic two-body interaction given by the Brueckner $G$-matrix based on the CD Bonn potential. But these works did not explicitly include the deformation, and the IS $np$ pairing was taken into account effectively by a renormalization of the IV $np$ pairing strength. Recent study \cite{Gamb15} regarding the relation of the deformation and the $np$ pairing correlations addressed extensively, by a model combining shell model techniques and mean-field, that the large coexistence of the IV and IS may be found and the deformation can induce the quenching of the IV $np$ pairing.

Furthermore, recent experimental data for the M1 spin transition
reported  strong IV quenching in the $N =Z$ $sd$-shell nuclei
\cite{Matsu15}, whose nuclei are thought to be largely deformed. It
means that the IV quenching giving rise to IS condensation
may become of an important ingredient for understanding the nuclear
deformation in those nuclei. In Refs. \cite{Ha17-2,Ha18}, we demonstrated
that such IS condensation is really feasible in those $sd$- and $pf$-shell
nuclei, in particular, for large deformation region.
Similar discussions were carried out by other authors \cite{Bai14,Nils04}. But the deformation was not explicitly taken into account. Ref. \cite{Bai14} argued
that the IS $np$ pairing may affect the low-lying Gamow-Teller (GT) state near to
the isobaric analogue resonance (IAR) for $pf$-shell $N=Z+2$ nuclei
by studying the GT data \cite{Fujita14}.
Ref. \cite{Nils04} has performed a self-consistent $pn$-QRPA based on a relativistic HFB approach, which  clearly demonstrated the importance of the $np$ pairing, specifically, IV $np$ pairing, for proper understanding the GT peaks. But, very recent calculation of the GT strength for Ni isotopes by $pn$-QRPA \cite{Sadiye18} is based only on the like-pairing correlations in the spherical QRPA.

Main interests in this work are how to interrelate the $np$ pairing in the residual interaction with the deformation considered in the mean field on the Gamow-Teller (GT) transition because the IS pairing may compete with the deformation features due to its coupling to non-spherical $J =$ odd states. Most studies have focused on the $N = Z$ nuclei because the $np$ pairing is expected to be larger rather than other $N \neq Z$ nuclei. However, as shown in a recent work \cite{Bertch11}, the nuclear structure of the $N \simeq Z$ nuclei may also be more or less affected by the $np$ pairing correlations.

In our recent papers \cite{Ha15,Ha15-1}, we developed a deformed QRPA (DQRPA) approach by explicitly including the deformation \cite{Ha15}, in which all effects by the deformation and the like-pairing ($nn$ and $pp$) correlations are consistently treated at the Deformed BCS (DBCS) and RPA stages. But the $np$ pairing correlations were taken into account only at the DBCS with a schematic interaction and a $G$-matrix type interaction \cite{Ha15-1}. Very recently both effects are simultaneously considered and applied to the GT strength study of $^{24,26}$Mg at Ref. \cite{Ha16} and other $N=Z$ $sd$-shell nuclei at Ref. \cite{Ha17}. We argued that the $np$ pairing correlations as well as the deformation may affect the $sd$-shell nuclear structure and their GT response functions. Along these preceding papers, here, we extend our applications to the GT strength study of $pf$-shell nuclei in order to understand the interplay of the pairing correlations and the deformation in the medium heavy nuclei.

In this work, we also investigate how such IS condensation and deformation affect the GT strength distribution in $pf$-shell nuclei, because roles of the IS and IV pairings in the deformed mean field and their effects on the GT strength distributions are still remained to be intensively discussed in those medium heavy nuclei. Our results are presented as follows. In section II, a brief pedagogical explanation of the formalism is introduced. Numerical GT results for $^{56,58}$N i and $^{62,64}$Ni are discussed in Sec. III. A summary and conclusions are done in Sec. IV.

\section{Theoretical Formalism}

The $np$ pairing
correlations change the conventional quasi-particle concept, {\it
i.e.}, quasi-neutron (quasi-proton) composed by neutron (proton) particle and its hole state, to quasi-particle 1 and 2
which may mix properties of the quasi-proton and
quasi-neutron. Here we explain briefly the formalism, DBCS and DQRPA, to be applied for GT transition strength distributions of some Ni isotopes, in which we include all types pairings as well as the deformation.
We start from the following DBCS transformation between creation and annihilation operators for real (bare) and quasi-particle states \cite{Ha15}
\begin{equation} \label{eq:HFB_1}
\left( \begin{array}{c} a_{1}^{\dagger} \\
  a_{2}^{\dagger} \\
  a_{\bar{1}} \\
  a_{\bar{2}}
  \end{array}\right)_{\alpha} =
\left( \begin{array}{cccc}
u_{1p} & u_{1n} & v_{1p} & v_{1n} \\
u_{2p} & u_{2n} & v_{2p} & v_{2n} \\
-v_{1p} & -v_{1n} & u_{1p} & u_{1n} \\
-v_{2p} & -v_{2n} & u_{2p} & u_{2n}
  \end{array}\right)_{\alpha}
\left( \begin{array}{c}
  c_{p}^{\dagger} \\
  c_{n}^{\dagger} \\
  c_{\bar{p}} \\
  c_{\bar{n}}
  \end{array}\right)_{\alpha} ~.
 \end{equation}
Hereafter Greek letter denotes a single-particle state (SPS) of neutron and proton with a projection $\Omega $ of a total angular momentum on
a nuclear symmetry axis. The projection $\Omega$ is treated as the only good quantum number in the deformed basis.
We assume the time reversal symmetry in the transformation coefficient and do not allow mixing of different SPSs ($\alpha$ and $\beta$) to a quasi-particle in the deformed state. 

But, in a spherical state representation of Eq. (\ref{eq:HFB_1}), the quasi-particle states would be mixed with different particle states in the spherical state because each deformed state (basis) is expanded by a linear combination of the spherical state (basis) (see Fig. 1 at Ref. \cite{Ha15}). In this respect, the DBCS is another representation of the deformed HFB transformation in the spherical basis. If we discard the $np$ pairing, Eq. (\ref{eq:HFB_1}) is reduced to the conventional BCS transformation in a deformed basis.

The other merit is that, by expanding all deformed wave functions constructed from the deformed harmonic oscillator basis into the spherical basis, we may exploit the Wigner-Eckardt theorem for more complicated physical operators in the deformed states. Finally, using the transformation of Eq. (\ref{eq:HFB_1}), the following DBCS equation
was obtained
\begin{equation} \label{eq:hfbeq}
\left( \begin{array}{cccc} \epsilon_{p}-\lambda_{p} & 0 &
\Delta_{p {\bar p}} & \Delta_{p {\bar n}} \\
0  & \epsilon_{n}-\lambda_{n} & \Delta_{n
{\bar p}} & \Delta_{n {\bar n}} \\
  \Delta_{p {\bar p}} &
 \Delta_{p {\bar n}} & -\epsilon_{p} + \lambda_{p} & 0 \\
  \Delta_{n {\bar p}} &
 \Delta_{n {\bar n}} & 0 & -\epsilon_{n} + \lambda_{n}
  \end{array}\right)_{\alpha}
\left( \begin{array}{c}
u_{\alpha'' p} \\ u_{\alpha'' n} \\ v_{\alpha'' p} \\
v_{\alpha'' n} \end{array}\right)_{\alpha}
 =
 E_{\alpha \alpha''}
\left( \begin{array}{c} u_{\alpha'' p} \\ u_{\alpha'' n} \\
 v_{\alpha'' p} \\
v_{\alpha'' n} \end{array}\right)_{\alpha},
\end{equation}
where $E_{\alpha \alpha''}$ is the energy of a quasi-particle denoted as $\alpha''$(=1 or 2) in the $\alpha$ state.
The pairing potentials in the DBCS Eq. (\ref{eq:hfbeq}) were calculated in the deformed basis by using the $G$-matrix obtained from the realistic Bonn CD potential for nucleon-nucleon (N-N) interaction as follows
\begin{equation} \label{eq:gap}
\Delta_{{p \bar{p}_\alpha}} = \Delta_{\alpha p \bar{\alpha}p} = -
\sum_{\gamma}  \Big[ \sum_{J, a, c } g_{\textrm{pp}} F_{\alpha a
\bar{\alpha} a}^{J0} F_{\gamma c \bar{\gamma} c}^{J0}
G(aacc,J,T=1)\Big] (u_{1p_{\gamma}}^* v_{1p_{\gamma}} +
u_{2p_{\gamma}}^* v_{2p_{\gamma}}) ~,
\end{equation}
\begin{eqnarray} \label{eq:gap_pn}
\Delta_{{p \bar{n}_\alpha}} = \Delta_{\alpha p \bar{\alpha}n} = &-&
\sum_{\gamma} \Bigg[ \Big[\sum_{J, a, c} g_{\textrm{np}} F_{\alpha a
\bar{\alpha} a}^{J0} F_{\gamma c \bar{\gamma} c}^{J0}
G(aacc,J,T=1)\Big] Re(u_{1n_{\gamma}}^* v_{1p_{\gamma}} +
u_{2n_{\gamma}}^* v_{2p_{\gamma}}) \\ \nonumber
 &+& \Big[ \sum_{J, a, c}
g_{\textrm{np}} F_{\alpha a \bar{\alpha} a}^{J0} F_{\gamma c
\bar{\gamma} c}^{J0} iG(aacc,J,T=0)\Big] Im (u_{1n_{\gamma}}^*
v_{1p_{\gamma}} + u_{2n_{\gamma}}^* v_{2p_{\gamma}}) \Bigg]~,
\end{eqnarray}
where $F_{ \alpha a  {\bar \alpha a
}}^{JK}=B_{a}^{\alpha}~B_{a}^{\alpha} ~{(-1)^{j_{a} -\Omega_{\alpha}}}~C^{JK}_{j_{a}
\Omega_{\alpha} j_{a}-\Omega_{\alpha}}(K=\Omega_{\alpha} - \Omega_{\alpha})$ was introduced to describe the $G$-matrix in the deformed basis with the expansion coefficient $B_{\alpha}$ calculated as \cite{Ha15}
\begin{equation} \label{eq:sps_ex}
B_{a}^{\alpha} = \sum_{N n_z \Sigma} C_{l \Lambda { 1 \over 2} \Sigma}^{j \Omega_{\alpha}}
A_{N n_z \Lambda}^{N_0 l}~b_{N n_z \Sigma} ~,~A_{N n_z \Lambda}^{N_0 l n_r} =<N_0 l \Lambda|N n_z \Lambda
>~.
\end{equation}
Here $K$ is a projection number of the total angular momentum $J$ onto the $z$ axis and selected as $K=0$ at the DBCS stage because we considered pairings of the quasi-particles at $\alpha$ and ${\bar\alpha}$ states. $G(aacc ~J T)$ represents a two-body (pairwise) scattering matrix in the spherical basis, where all possible scattering of the nucleon pairs above Fermi surface were taken into account.

In the present work, we have included all possible $J$ values with the $K =0$ projection.
$\Delta_{\alpha n \bar{\alpha}n}$ is the similar to Eq. (\ref{eq:gap}) where $n$ was replaced by $p$.
In order to renormalize the $G$-matrix due to the finite Hilbert space, strength parameters,
$g_{\textrm{pp}}$, $g_{\textrm{nn}}$, and $g_{\textrm{np}}$ were multiplied with the $G$-matrix \cite{Ch93} by adjusting the pairing potentials, $\Delta_{p {\bar p}}$ and $\Delta_{n {\bar n}}$, in Eq. (\ref{eq:gap}) to the empirical pairing gaps, $\Delta_{p}^{emp}$ and $\Delta_{n}^{emp}$, which were evaluated by a symmetric five mass term formula for the neighboring nuclei \cite{Ha15-1} using empirical masses.
The theoretical $np$ pairing gaps were calculated as \cite{Ch93,Bend00}
\begin{equation}
\delta_{np}^{th.} = - [ ( H_{gs}^{12} + E_1 + E_2 ) - ( H_{gs}^{np} + E_p + E_n)]~.
\end{equation}
Here $H_{gs}^{12} (H_{gs}^{np}) $ is a total DBCS ground state energy with (without) $np$ pairing and $ E_1 + E_2 (E_p + E_n)$ is a sum of the lowest two quasi-particle energies with (without) $np$ pairing potential $\Delta_{n{\bar p}}$ in Eq. (\ref{eq:hfbeq}).

For the mean field energy $\epsilon_{p(n)}$ in Eq. (\ref{eq:hfbeq}) we exploited a deformed Woods-Saxon potential
\cite{cwi} with the universal parameter set. By taking the same approach as used in the QRPA equation
in Ref. \cite{Suhonen}, our Deformed QRPA (DQRPA) equation including the $np$ pairing
correlations was summarized as follows,
\begin{eqnarray}\label{qrpaeq}
&&\left( \begin{array}{cccccccc}
          ~ A_{\alpha \beta \gamma \delta}^{1111}(K) &~ A_{\alpha \beta \gamma \delta}^{1122}(K) &
          ~ A_{\alpha \beta \gamma \delta}^{1112}(K) &~ A_{\alpha \beta \gamma \delta}^{1121}(K) &
          ~ B_{\alpha \beta \gamma \delta}^{1111}(K) &~ B_{\alpha \beta \gamma \delta}^{1122}(K) &
          ~ B_{\alpha \beta \gamma \delta}^{1112}(K) &~ B_{\alpha \beta \gamma \delta}^{1121}(K) \\
          ~ A_{\alpha \beta \gamma \delta}^{2211}(K) &~ A_{\alpha \beta \gamma \delta}^{2222}(K) &
          ~ A_{\alpha \beta \gamma \delta}^{2212}(K) &~ A_{\alpha \beta \gamma \delta}^{2221}(K) &
           ~B_{\alpha \beta \gamma \delta}^{2211}(K) & ~B_{\alpha \beta \gamma \delta}^{2222}(K) &
           ~B_{\alpha \beta \gamma \delta}^{2212}(K) & ~B_{\alpha \beta \gamma \delta}^{2221}(K)\\
           ~A_{\alpha \beta \gamma \delta}^{1211}(K) & ~A_{\alpha \beta \gamma \delta}^{1222}(K) &
           ~A_{\alpha \beta \gamma \delta}^{1212}(K) & ~A_{\alpha \beta \gamma \delta}^{1221}(K) &
           ~B_{\alpha \beta \gamma \delta}^{1211}(K) & ~B_{\alpha \beta \gamma \delta}^{1222}(K) &
           ~B_{\alpha \beta \gamma \delta}^{1212}(K) & ~B_{\alpha \beta \gamma \delta}^{1221}(K)\\
           ~A_{\alpha \beta \gamma \delta}^{2111}(K) & ~A_{\alpha \beta \gamma \delta}^{2122}(K) &
           ~A_{\alpha \beta \gamma \delta}^{2112}(K) & ~A_{\alpha \beta \gamma \delta}^{2121}(K) &
           ~B_{\alpha \beta \gamma \delta}^{2111}(K) & ~B_{\alpha \beta \gamma \delta}^{2122}(K) &
           ~B_{\alpha \beta \gamma \delta}^{2112}(K) & ~B_{\alpha \beta \gamma \delta}^{2121}(K)\\
             &       &       &      &      &        &           &        \\
           - B_{\alpha \beta \gamma \delta}^{1111}(K) & -B_{\alpha \beta \gamma \delta}^{1122}(K) &
            -B_{\alpha \beta \gamma \delta}^{1112}(K) & -B_{\alpha \beta \gamma \delta}^{1121}(K) &
           - A_{\alpha \beta \gamma \delta}^{1111}(K) & -A_{\alpha \beta \gamma \delta}^{1122}(K) &
           -A_{\alpha \beta \gamma \delta}^{1112}(K)  & -A_{\alpha \beta \gamma \delta}^{1121}(K)\\
           - B_{\alpha \beta \gamma \delta}^{2211}(K) & -B_{\alpha \beta \gamma \delta}^{2222}(K) &
           -B_{\alpha \beta \gamma \delta}^{2212}(K)  & -B_{\alpha \beta \gamma \delta}^{2221}(K) &
           - A_{\alpha \beta \gamma \delta}^{2211}(K) & -A_{\alpha \beta \gamma \delta}^{2222}(K) &
           -A_{\alpha \beta \gamma \delta}^{2212}(K)  & -A_{\alpha \beta \gamma \delta}^{2221}(K)\\
           - B_{\alpha \beta \gamma \delta}^{1211}(K) & -B_{\alpha \beta \gamma \delta}^{1222}(K) &
           -B_{\alpha \beta \gamma \delta}^{1212}(K)  & -B_{\alpha \beta \gamma \delta}^{1221}(K) &
           - A_{\alpha \beta \gamma \delta}^{1211}(K) & -A_{\alpha \beta \gamma \delta}^{1222}(K) &
           -A_{\alpha \beta \gamma \delta}^{1212}(K)  & -A_{\alpha \beta \gamma \delta}^{1221}(K) \\
          - B_{\alpha \beta \gamma \delta}^{2111}(K) & -B_{\alpha \beta \gamma \delta}^{2122}(K) &
           -B_{\alpha \beta \gamma \delta}^{2112}(K)  & -B_{\alpha \beta \gamma \delta}^{2121}(K) &
           - A_{\alpha \beta \gamma \delta}^{2111}(K) & -A_{\alpha \beta \gamma \delta}^{2122}(K) &
           -A_{\alpha \beta \gamma \delta}^{2112}(K)  & -A_{\alpha \beta \gamma \delta}^{2121}(K) \\
           \end{array} \right)\\ \nonumber ~~&& \times
\left( \begin{array}{c}   {\tilde X}_{(\gamma 1 \delta 1)K}^{m}  \\ {\tilde X}_{(\gamma 2 \delta 2)K}^{m} \\
  {\tilde X}_{(\gamma 1 \delta 2)K}^{m} \\  {\tilde X}_{(\gamma 2 \delta 1)K}^{m} \\ \\
     {\tilde Y}_{(\gamma 1 \delta 1)K}^{m} \\ {\tilde Y}_{(\gamma 2 \delta 2)K}^{m} \\
     {\tilde Y}_{(\gamma 1 \delta 2)K}^{m}\\{\tilde Y}_{(\gamma 2 \delta 1)K}^{m} \end{array} \right)
 = \hbar {\Omega}_K^{m}
 \left ( \begin{array}{c} {\tilde X}_{(\alpha 1 \beta 1)K}^{m}  \\{\tilde X}_{(\alpha 2 \beta 2)K}^{m} \\
 {\tilde X}_{(\alpha 1 \beta 2)K}^{m} \\  {\tilde X}_{(\alpha 2 \beta 1)K}^{m}\\ \\
{\tilde Y}_{(\alpha 1 \beta 1)K}^{m} \\ {\tilde Y}_{(\alpha 2 \beta 2)K}^{m} \\
{\tilde Y}_{(\alpha 1 \beta 2)K}^{m} \\ {\tilde Y}_{(\alpha 2 \beta 1)K}^{m} \end{array} \right)  ~,
\end{eqnarray}
where the amplitudes
${\tilde X}^m_{(\alpha \alpha''  \beta \beta'')K }$ and ${\tilde Y}^m_{(\alpha
\alpha''  \beta \beta'')K}$ in Eq. (\ref{qrpaeq}) stand for forward and backward going amplitudes from state ${ \alpha
\alpha'' }$ to ${\beta  \beta''}$ state \cite{Ch93}.

Our DQRPA equation is very general because we include the deformation as well as all kinds of pairing correlations still remained in the mean field.
If we switch off the $np$ pairing, all off-diagonal terms in the A and B matrices in Eq. (\ref{qrpaeq}) disappear with the replacement of 1 and 2 into $p$ and $n$. Then the DQRPA equation is decoupled into pp + nn + pn + np DQRPA equations \cite{saleh}. For charge conserving (or neutral current) reactions, pp + nn DQRPA can describe the M1 spin or EM transitions on the same nuclear species, while np + pn DQRPA describes the GT(+/-) transitions in the charge exchange (or charged current) reactions. Here it should be pointed out that np DQRPA is different from pn DQRPA because of the deformation. The A and B matrices in Eq. (\ref{qrpaeq}) are given by
\begin{eqnarray} \label{eq:mat_A}
A_{\alpha \beta \gamma \delta}^{\alpha'' \beta'' \gamma'' \delta''}(K)  = && (E_{\alpha
   \alpha''} + E_{\beta \beta''}) \delta_{\alpha \gamma} \delta_{\alpha'' \gamma''}
   \delta_{\beta \delta} \delta_{\beta'' \delta''}
       - \sigma_{\alpha \alpha'' \beta \beta''}\sigma_{\gamma \gamma'' \delta \delta''}\\ \nonumber
   &\times&
   \sum_{\alpha' \beta' \gamma' \delta'}
   [-g_{pp} (u_{\alpha \alpha''\alpha'} u_{\beta \beta''\beta'} u_{\gamma \gamma''\gamma'} u_{\delta \delta''\delta'}
   +v_{\alpha \alpha''\alpha'} v_{\beta \beta''\beta'} v_{\gamma \gamma''\gamma'} v_{\delta \delta''\delta'} )
    ~V_{\alpha \alpha' \beta \beta',~\gamma \gamma' \delta \delta'}
    \\ \nonumber  &-& g_{ph} (u_{\alpha \alpha''\alpha'} v_{\beta \beta''\beta'}u_{\gamma \gamma''\gamma'}
     v_{\delta \delta''\delta'}
    +v_{\alpha \alpha''\alpha'} u_{\beta \beta''\beta'}v_{\gamma \gamma''\gamma'} u_{\delta \delta''\delta'})
    ~V_{\alpha \alpha' \delta \delta',~\gamma \gamma' \beta \beta'}
     \\ \nonumber  &-& g_{ph} (u_{\alpha \alpha''\alpha'} v_{\beta \beta''\beta'}v_{\gamma \gamma''\gamma'}
     u_{\delta \delta''\delta'}
     +v_{\alpha \alpha''\alpha'} u_{\beta \beta''\beta'}u_{\gamma \gamma''\gamma'} v_{\delta \delta''\delta'})
    ~V_{\alpha \alpha' \gamma \gamma',~\delta \delta' \beta \beta' }],
\end{eqnarray}
\begin{eqnarray} \label{eq:mat_B}
B_{\alpha \beta \gamma \delta}^{\alpha'' \beta'' \gamma'' \delta''}(K)  =
 &-& \sigma_{\alpha \alpha'' \beta \beta''} \sigma_{\gamma \gamma'' \delta \delta''}
  \\ \nonumber &\times&
 \sum_{\alpha' \beta' \gamma' \delta'}
  [g_{pp}
  (u_{\alpha \alpha''\alpha'} u_{\beta \beta''\beta'}v_{\gamma \gamma''\gamma'} v_{\delta \delta''\delta'}
   +v_{\alpha \alpha''\alpha'} v_{{\bar\beta} \beta''\beta'}u_{\gamma \gamma''\gamma'} u_{{\bar\delta} \delta''\delta'} )
   ~ V_{\alpha \alpha' \beta \beta',~\gamma \gamma' \delta \delta'}\\ \nonumber
     &- & g_{ph} (u_{\alpha \alpha''\alpha'} v_{\beta \beta''\beta'}v_{\gamma \gamma''\gamma'}
     u_{\delta \delta''\delta'}
    +v_{\alpha \alpha''\alpha'} u_{\beta \beta''\beta'}u_{\gamma \gamma''\gamma'} v_{\delta \delta''\delta'})
   ~ V_{\alpha \alpha' \delta \delta',~\gamma \gamma' \beta \beta'}
     \\ \nonumber  &- & g_{ph} (u_{\alpha \alpha''\alpha'} v_{\beta \beta''\beta'}u_{\gamma \gamma''\gamma'}
      v_{\delta \delta''\delta'}
     +v_{\alpha \alpha''\alpha'} u_{\beta \beta''\beta'}v_{\gamma \gamma''\gamma'} u_{\delta \delta''\delta'})
   ~ V_{\alpha \alpha' \gamma \gamma',~\delta \delta' \beta \beta'}],
\end{eqnarray}
where $u$ and $v$ coefficients are determined from DBCS Eq. (\ref{eq:hfbeq}). The two body interactions $V_{\alpha \beta,~\gamma \delta}$ and $V_{\alpha \delta,~\gamma \beta}$
are particle-particle and particle-hole matrix elements of the residual $N-N$ interaction $V$, respectively, in the deformed state, which are detailed at Ref. \cite{Ha15-1}.

The GT transition amplitudes from the ground state of an initial (parent) nucleus
to the excited state of a daughter nucleus, {\it i.e.} the one phonon state
$K^+$ in a final nucleus, are written as
\begin{eqnarray} \label{eq:phonon}
&&< K^+, m | {\hat {GT}}_{K }^- | ~QRPA >  \\ \nonumber
&&= \sum_{\alpha \alpha''\rho_{\alpha} \beta \beta''\rho_{\beta}}{\cal N}_{\alpha \alpha''\rho_{\alpha}
 \beta \beta''\rho_{\beta} }
 < \alpha \alpha''p \rho_{\alpha}|  \sigma_K | \beta \beta''n \rho_{\beta}>
 [ u_{\alpha \alpha'' p} v_{\beta \beta'' n} X_{(\alpha \alpha''\beta \beta'')K}^{m} +
v_{\alpha \alpha'' p} u_{\beta \beta'' n} Y_{(\alpha \alpha'' \beta \beta'')K}^{m}], \\ \nonumber
&&< K^+, m | {\hat {GT}}_{K }^+ | ~QRPA >  \\ \nonumber
&&= \sum_{\alpha \alpha'' \rho_{\alpha} \beta \beta''\rho_{\beta}}{\cal N}_{\alpha \alpha'' \beta \beta'' }
 < \alpha \alpha''p \rho_{\alpha}|  \sigma_K | \beta \beta''n \rho_{\beta}>
 [ u_{\alpha \alpha'' p} v_{\beta \beta'' n} Y_{(\alpha \alpha'' \beta \beta'')K}^{m} +
v_{\alpha \alpha'' p} u_{\beta \beta'' n} X_{(\alpha \alpha'' \beta \beta'')K}^{m} ]~,
\end{eqnarray}
where $|~QRPA >$ denotes the correlated QRPA ground state in the intrinsic frame and
the nomalization factor is given as $ {\cal N}_{\alpha \alpha'' \beta
 \beta''} (J) = \sqrt{ 1 - \delta_{\alpha \beta} \delta_{\alpha'' \beta''} (-1)^{J + T} }/
 (1 + \delta_{\alpha \beta} \delta_{\alpha'' \beta''}). $ The forward and backward amplitudes, $X^m_{(\alpha \alpha''
 \beta \beta'')K}$ and $Y^m_{(\alpha \alpha''
 \beta \beta'')K}$, are related to
$\tilde{X^m}_{(\alpha \alpha'' \beta \beta'')K}=\sqrt2 \sigma_{\alpha \alpha'' \beta \beta''} X^m_{(\alpha \alpha''
 \beta \beta'')K}$
and $\tilde{Y^m}_{(\alpha \alpha'' \beta \beta'')K}=\sqrt2 \sigma_{\alpha \alpha'' \beta \beta''}
Y^m_{(\alpha \alpha'' \beta \beta'')K}$ in Eq. (\ref{qrpaeq}), where $\sigma_{\alpha \alpha'' \beta \beta''}$ = 1 if $\alpha = \beta$ and $\alpha''$ =
$\beta''$, otherwise $\sigma_{\alpha \alpha'' \beta \beta'' }$ = $\sqrt 2$ \cite{Ch93,Ha17}.

The particle model space
for all Ni isotopes was exploited up to $N = 5 \hbar \omega$ for a
deformed basis and up to $N = 10 \hbar \omega$ for a spherical
basis. In our previous papers \cite{Ha15,Ha15-1,Ha2013}, SPSs obtained from
the deformed Woods-Saxon potential were shown to be sensitive to the
deformation parameter $\beta_2$, which causes the shell evolution by the deformation. In this work, we allow some small variation within 20 \% from theoretical $\beta_2$ values in Table I to test the deformation dependence.

%
\section{Numerical results for the GT transition strength distribution for $^{56,58}$Ni and $^{62,64}$Ni}

\begin{table}
\caption[bb]{ Deformation parameter $\beta_2^{E2}$ from the
experimental E2 transition data \cite{Ram01} and theoretical
calculations, $\beta_2^{RMF}$ and $\beta_2^{FRDM}$, by Relativistic Mean Field (RMF) \cite{Lala99} and FRDM
model \cite{Mol16} for some Ni isotopes. Empirical pairing gaps evaluated from the five-point mass formula \cite{Ch93} are also shown.
Last column denotes Ikeda sum rule for GT transition \cite{Ha15} as a percentage ratio to $3(N-Z)$.
\\} \setlength{\tabcolsep}{2.0 mm}
\begin{tabular}{cccccccc}\hline
 Nucleus & $|\beta_2^{E2}|$ & $\beta_2^{RMF}$&$\beta_2^{FRDM}$ & $\Delta_p^{\textrm{emp}}$  &  $\Delta_n^{\textrm{emp}}$  & $\delta_{np}^{\textrm{emp}}$ & {GT($\%$)}
  \\ \hline \hline
$^{56}$Ni & 0.173 &0.     & 0.   & 2.077   & 2.150 & 1.107  & 99.6      \\
$^{58}$Ni & 0.182 &-0.001 & 0.   & 1.669   & 1.349 & 0.233  & 98.6    \\
$^{62}$Ni & 0.197 & 0.093 &0.107 & 1.747   & 1.639 & 0.465  & 99.1   \\
$^{64}$Ni & 0.179 &-0.091 &-0.094&1.747    & 1.602 & 0.454  & 99.2
\\ \hline
 \end{tabular}
\label{tab:beta2}
\end{table}

For direct comprehension of feasible effects due to the $np$ pairing and the
deformation in the GT transition, first, we took two Ni isotopes,
$^{56,58}$Ni, which are known as the nuclei affected easily by the $np$
pairing correlations \cite{Saga16,Bai13}. $^{56}$Ni is thought to be almost spherical because of its double magic number. But, in this work, we allowed small deformation, from spherical ($\beta_2$ = 0.0) to small deformation ($\beta_2$ = 0.15), in order to study the correlations of the deformation and the $np$ pairing. We note that if we take a $\alpha$-cluster model for $^{56}$Ni, the ground state may be deformed \cite{Darai11}. Second, we calculated the GT strength distribution for $^{62,64}$Ni, which have more excess neutrons with finite deformation. Moreover, they have stronger $np$ pairing gaps, almost twice, rather than $^{58}$Ni as shown in Table I, where we show the
empirical pairing gaps, the deformation parameter $\beta_{2}$
deduced from E2 transition data and theoretical estimations for Ni isotopes. Also we show Ikeda sum rule results for GT strength distribution.

%
\begin{figure}
\includegraphics[width=0.45\linewidth]{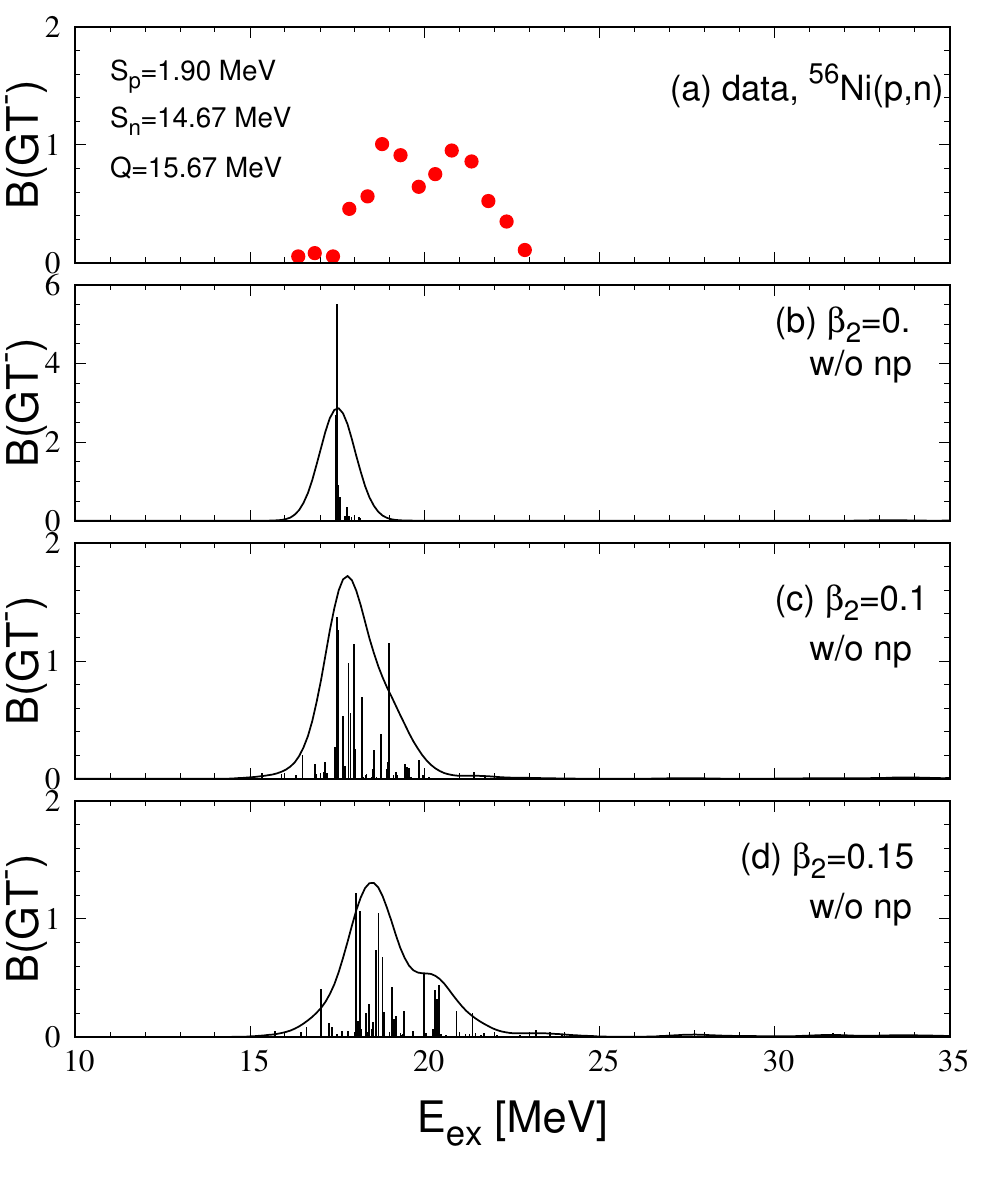}
\includegraphics[width=0.45\linewidth]{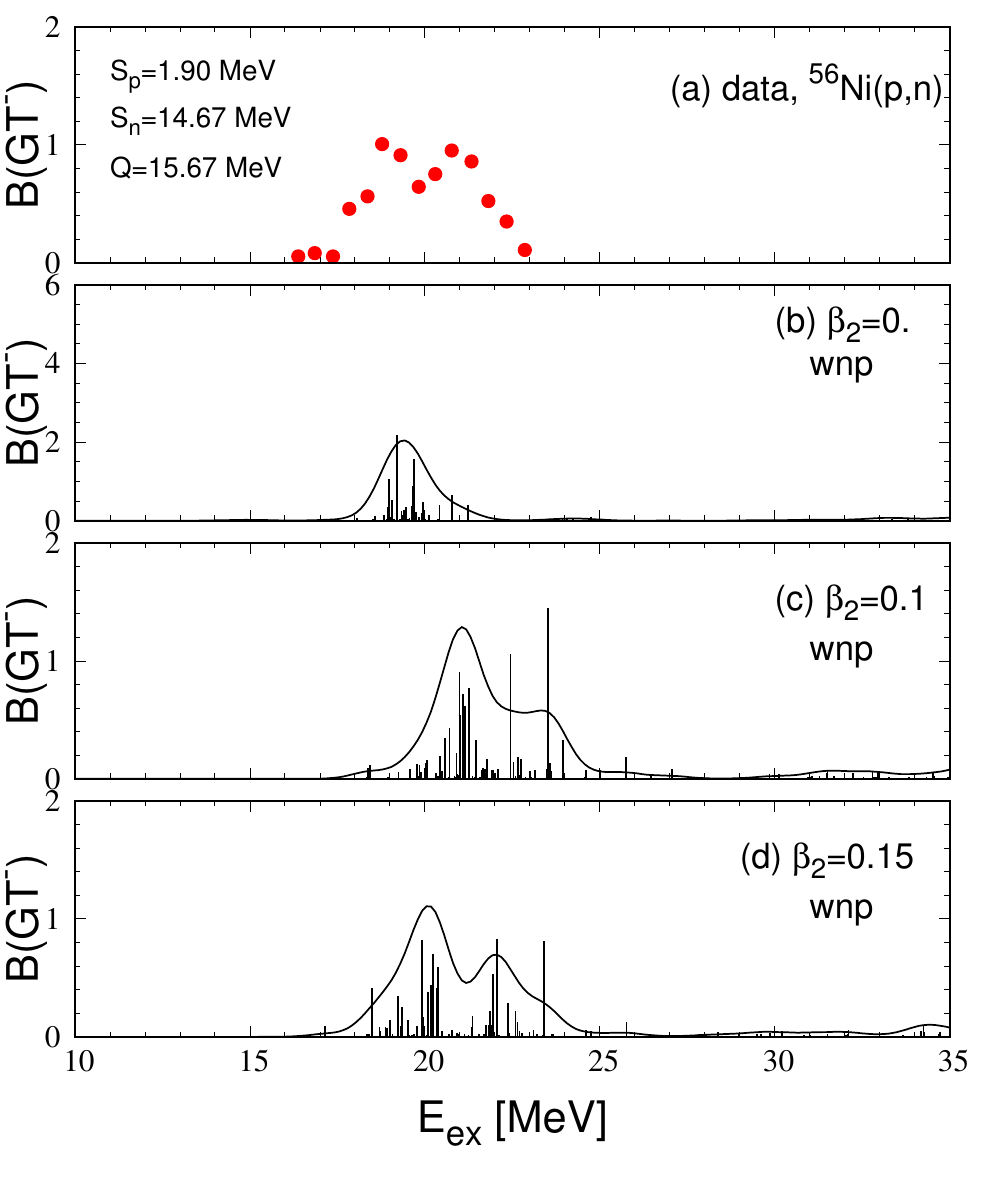}
\caption{(Color online) Gamow-Teller (GT) transition strength
distributions B(GT(--)) of $^{56}$Ni.
 Experimental data by $^{56}$Ni(p,n) in panel (a) are from Ref. \cite{Sasano11}.
 Results of (b) - (d) in the left (right) hand side are without (with) the $np$ pairing.
 Our results are presented by the excitation energy from parent nucleus.
 } \label{fig1}
\end{figure}

Recent calculations by the QRPA with particle-vibration coupling (PVC) based on the Skyrme interaction \cite{Niu14} addressed that the PVC contribution may spread or redistribute the GT strength for the double magic nucleus and also other Ni isotopes. But, if we recollect that the PVC contribution originates from the particle or hole propagation inside nuclei, these contributions can be taken into account by the Brueckner $G$-matrix in the Brueckner HF (BHF) approach, which already includes such nucleon-nucleon interaction inside nuclei through the Bethe-Goldstone equation. In the following, we discuss our numerical results for the GT strength distributions for $^{56,58}$Ni and $^{62,64}$Ni.

Figure \ref{fig1} presents GT strength distributions for
$^{56}$Ni(p,n) reaction by our DQRPA. Left (right) panels are
results without (with) the $np$
pairing correlations. In the left panel, the more deformation scatters the
distribution to the bit higher energy regions because of the repulsive particle-hole ($p-h$) interaction. But, the two peaks
peculiar to this GT distribution data were not reproduced well only by
the deformation. Namely, the 2nd high energy peak does not appear enough to explain the data.

In the right panel, we showed the $np$ pairing effects, which push the
distribution to the higher energy region even without the deformation, if one compares  the results in the left panel to the right panel with the same deformation.
Contrary to the $p-h$ repulsive force by the deformation, the $np$ pairing is mainly attractive, by which the Fermi energy difference of protons and neutrons, $\Delta_f = \epsilon_f^p - \epsilon_f^n$, is reduced by its attractive interaction and consequently gives rise to high-lying GT transitions between deeply bound neutron and proton single particle states \cite{Ha17}. As a result, the two peaks and their magnitudes appear explicitly by
the $np$ pairing.
\begin{figure}
\includegraphics[width=0.6\linewidth]{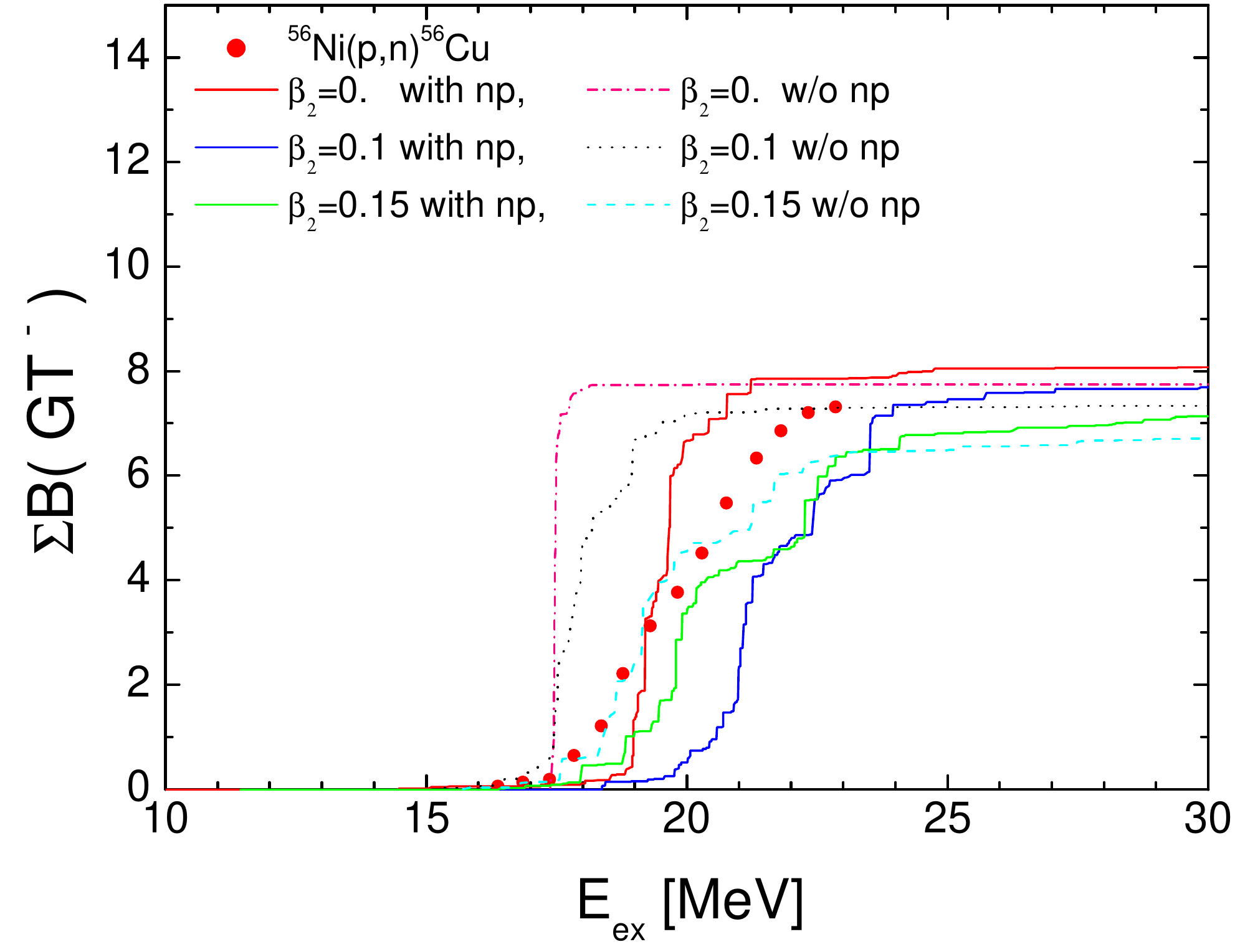}
\caption{(Color online)  Running sums for the GT (--) strength
distributions in Fig.\ref{fig1} (b)-(d) for $^{56}$Ni.
 } \label{fig2}
\end{figure}

This feature becomes significant in the running sum in Fig. \ref{fig2}, if one notes differences of the solid lines (with the $np$) and the dashed lines (without the $np$). Therefore, the deformation just scatters the strength distributions to a higher energy region by the repulsive $p-h$ interaction, but the $np$ pairing shifts the distribution to a higher energy region in a more concentrated form owing to the attractive interaction.

\begin{figure}
\includegraphics[width=0.65\linewidth]{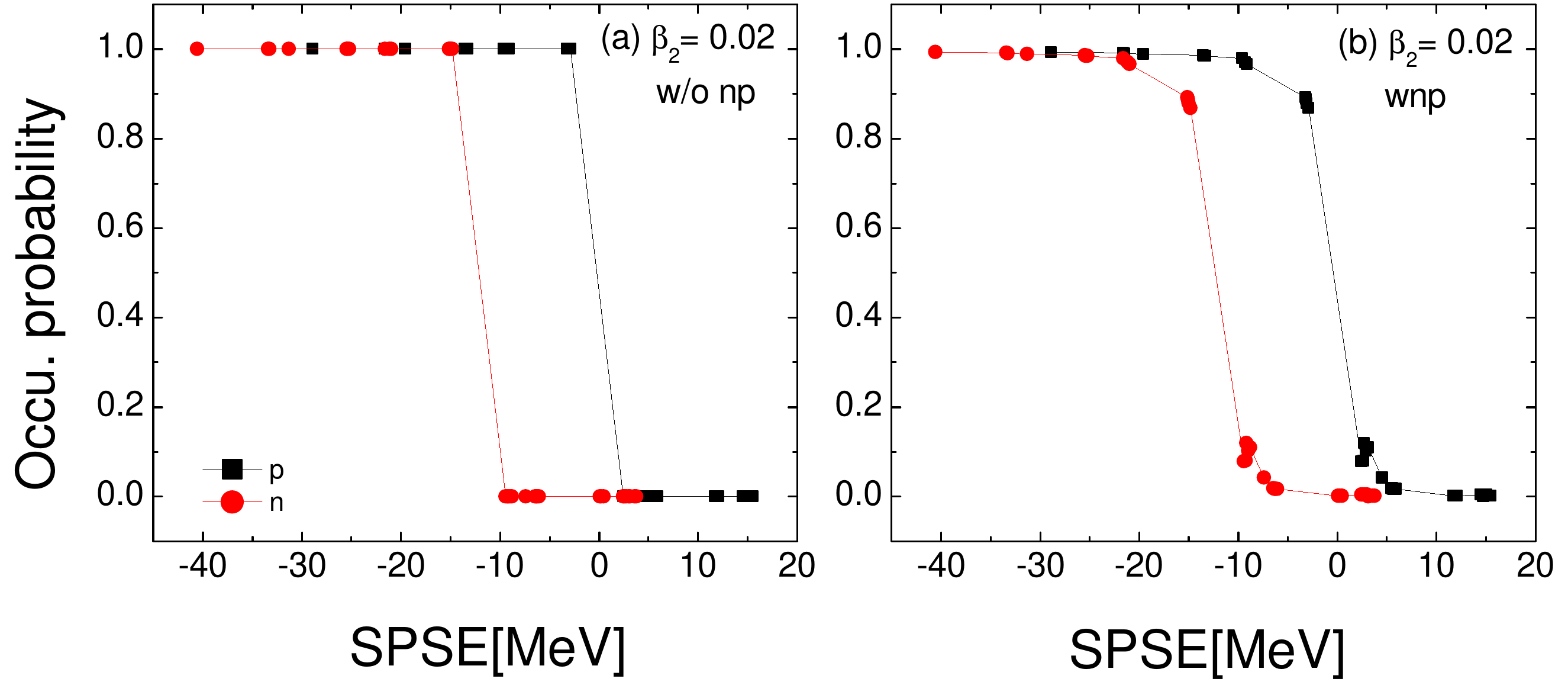}
\includegraphics[width=0.65\linewidth]{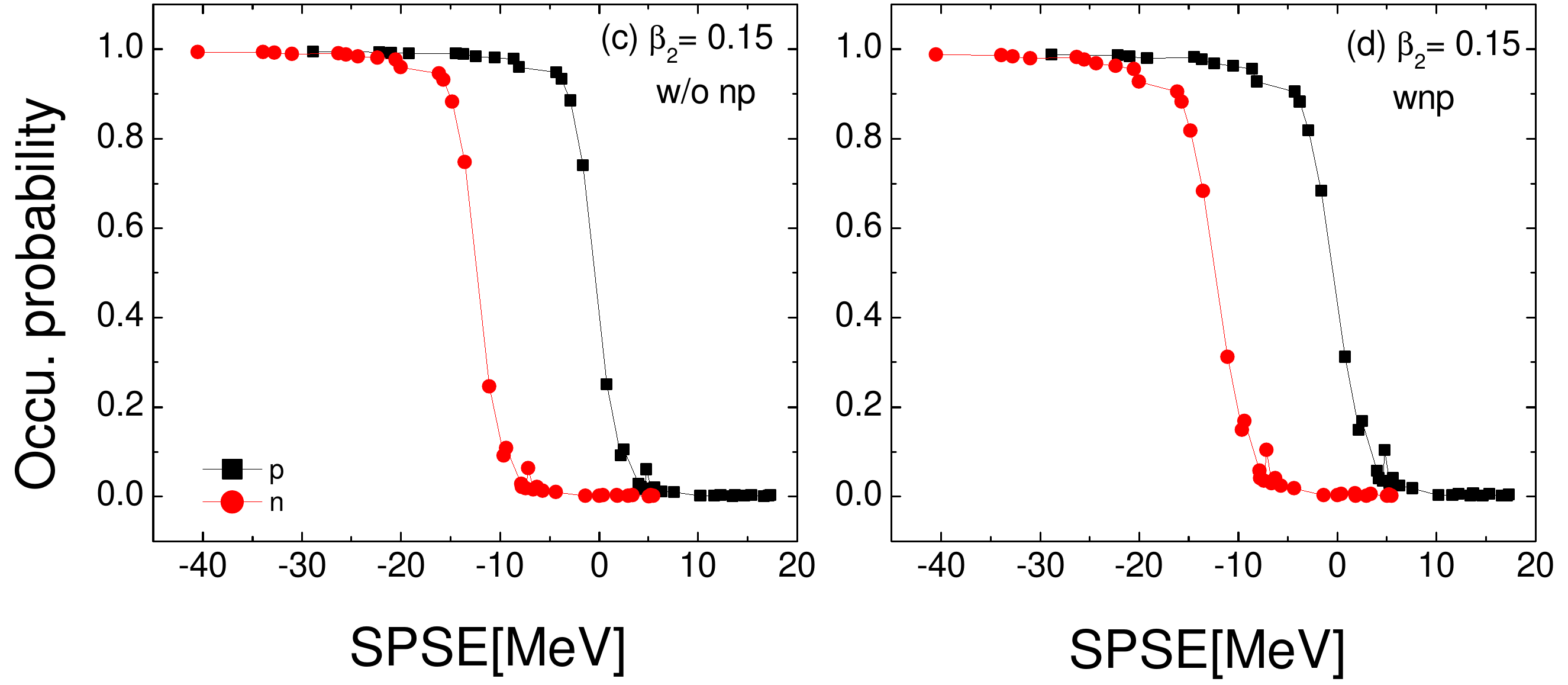}
\caption{(Color online) Occupation probabilities of neutrons and
protons in $^{56}$Ni of the single particle state energy
(SPSE) given by Nilsson basis \cite{Nilsson}. Left (right) panels are
with (without) neutron-proton pairing for $\beta_2=0.02$ ((a) and (b)) and $0.15$ ((c) and (d)), respectively.
 } \label{fig3}
\end{figure}
\begin{figure}
\includegraphics[width=0.45\linewidth]{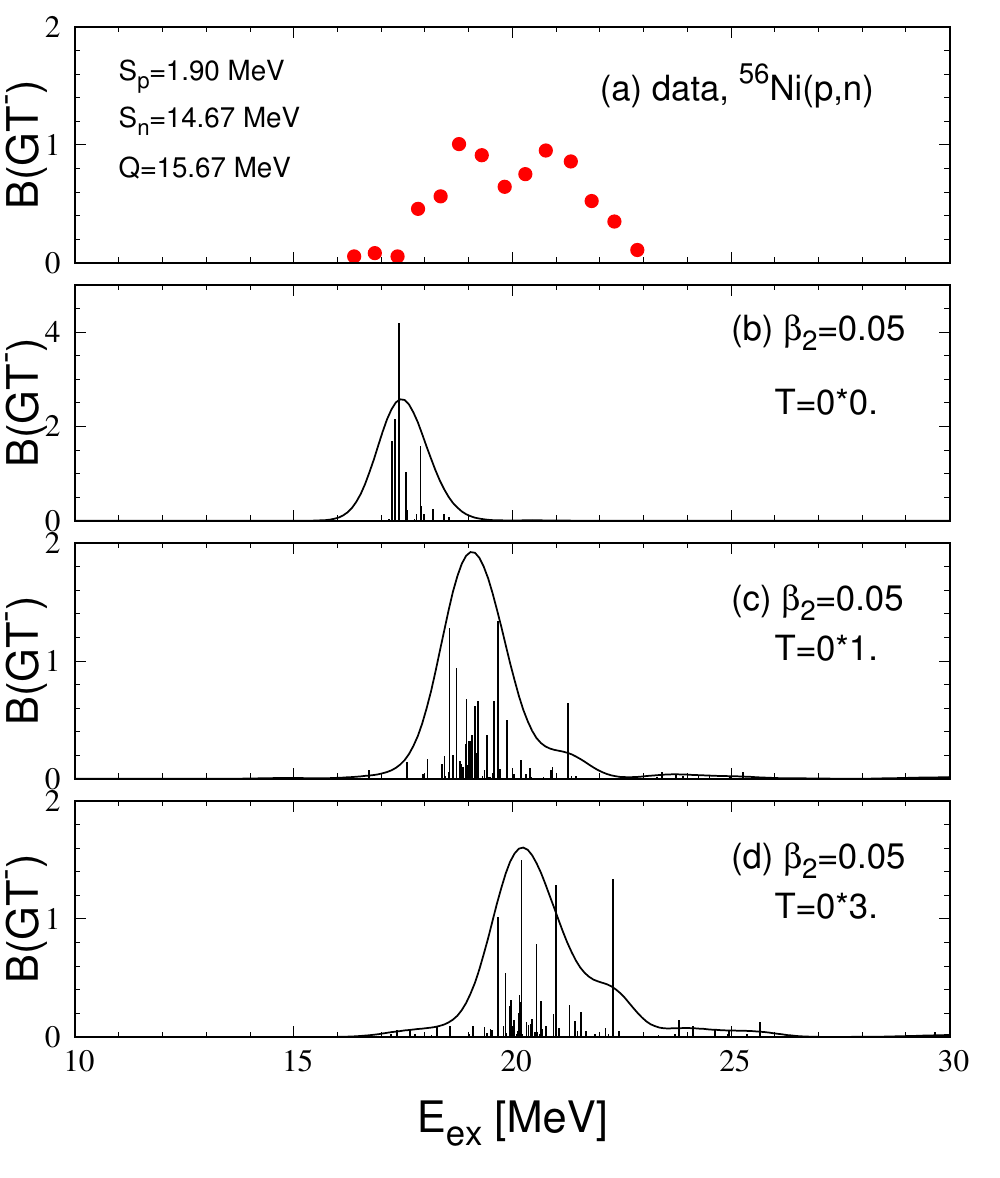}
\includegraphics[width=0.45\linewidth]{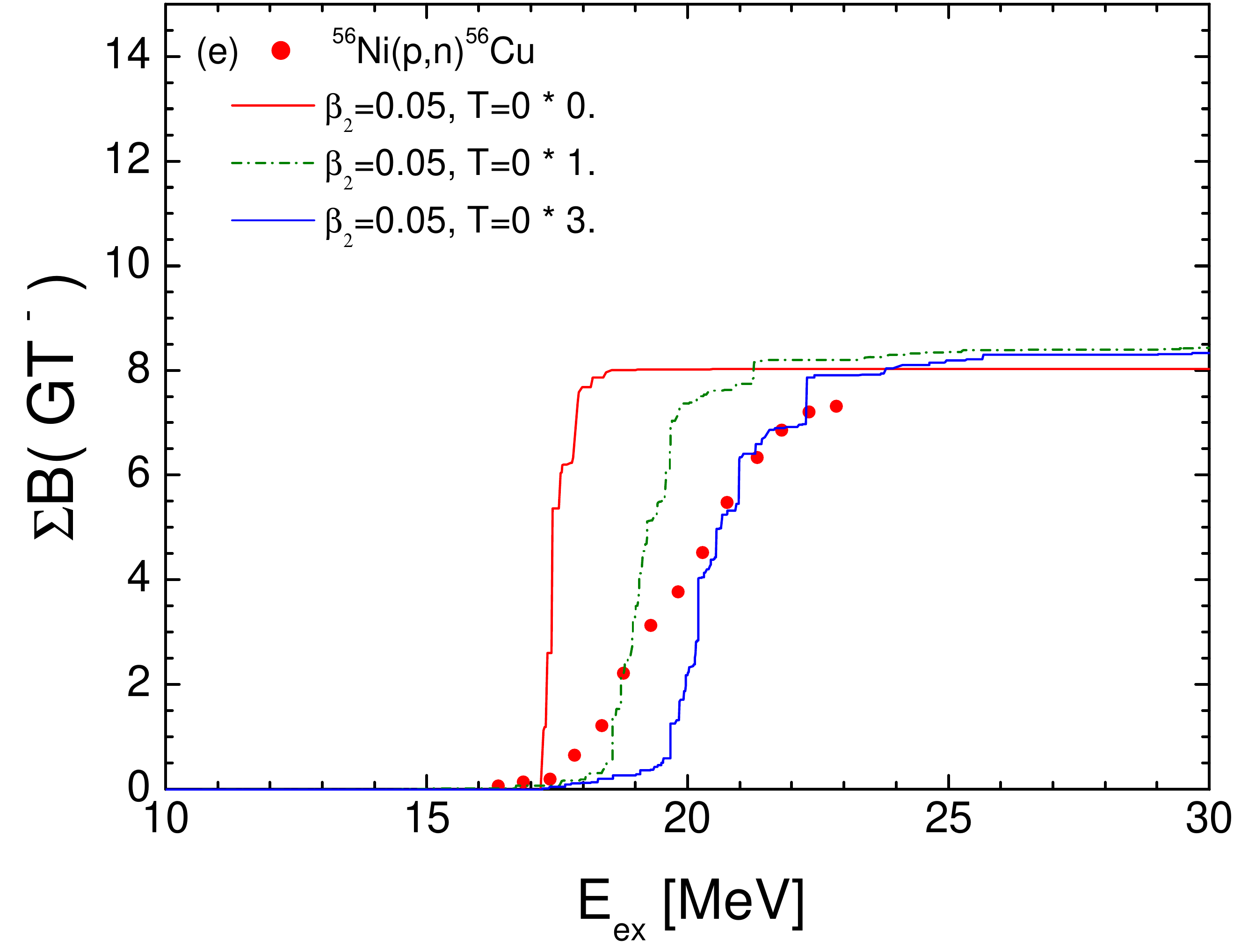}
\caption{(Color online) Isoscalar (IS) $np$ pairing effects on the GT (--) strength distribution (left) and its running sum (right) for $^{56}$Ni with $\beta_2$ = 0.05. Panels in the left side show results without the IS $np$ pairing (a) ($T=0*0$), with the normal IS $np$ pairing (b) ($T =0*1$)  and with the enhanced IS $np$ pairing (c) ($T =0*3$), respectively. Panel (e) is their running sums.} \label{fig4}
\end{figure}

Figure \ref{fig3}  shows change of the Fermi surface by the deformation and the $np$ pairing correlations. The larger deformation makes the more smearing as shown Fig.\ref{fig3}  (c) and (d). But the $np$ pairing gives rise to more effective smearing, which leads to the deeper Fermi energies, compared to the deformation effect, if we compare results in Fig.\ref{fig3} (a) and (c) to those in Fig.\ref{fig3} (b) and (d).

Recently, the IV quenching was reported at the $M1$ spin strength distribution of $N =Z$ $sd$-shell nuclei \cite{Matsu15}. Because this quenching may mean the condensation of the IS pairing in the $np$ pairing, we tested the IS pairing effects on the GT distribution \cite{Ha17-2,Ha18}. Figure \ref{fig4} reveals the effects of the enhanced IS condensation with a small deformation $\beta_2 =$ 0.05. The left panels explicitly show the shift of the GT strength distribution to a higher energy region with the increase of IS pairing, {\it i.e.} the case without IS (a), normal IS (c) used in the results of Fig. \ref{fig1} and the enhanced IS (d), where we retain the IV pairing to ensure the IS effect.

We took the enhanced IS pairing factor as 3.0 as argued in our previous paper \cite{Ha17-2,Ha18}. Because the IS pairing causes more attractive interaction, as shown in $^3 S_1$ state in the $np$ interaction, rather than the IV coupling, the shift of the GT strength distributions by the $np$ pairing is mainly attributed to the IS coupling condensation. This trend is also found in the results of Fig. \ref{fig1}. The IS and IV $np$ effects also manifest themselves in the GT running sum of Fig.\ref{fig4} (e). Not only the IV effect but also the IS effect are shown to be salient on the GT strength in this $N=Z$ nucleus.

\begin{figure}
\includegraphics[width=0.45\linewidth]{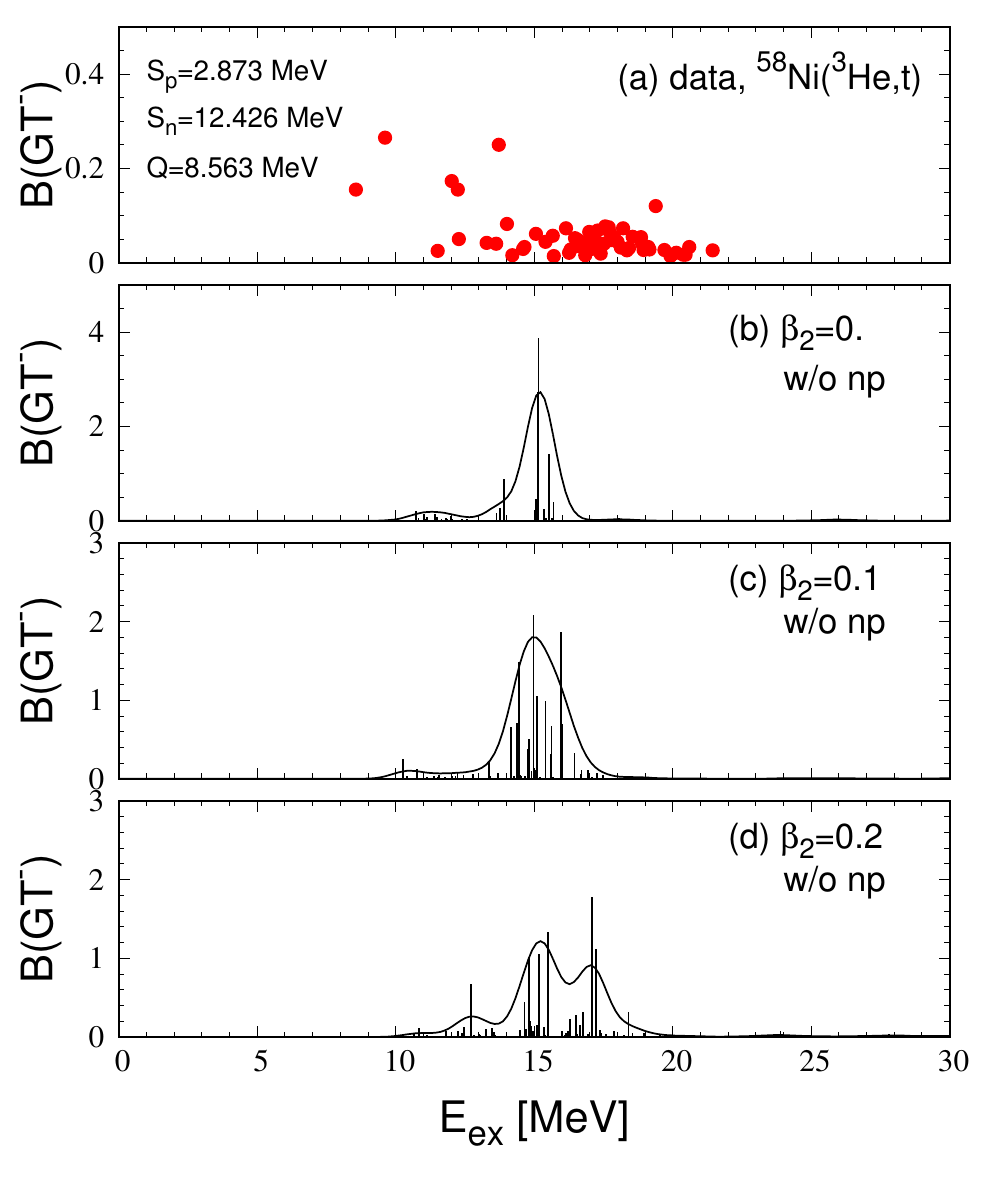}
\includegraphics[width=0.45\linewidth]{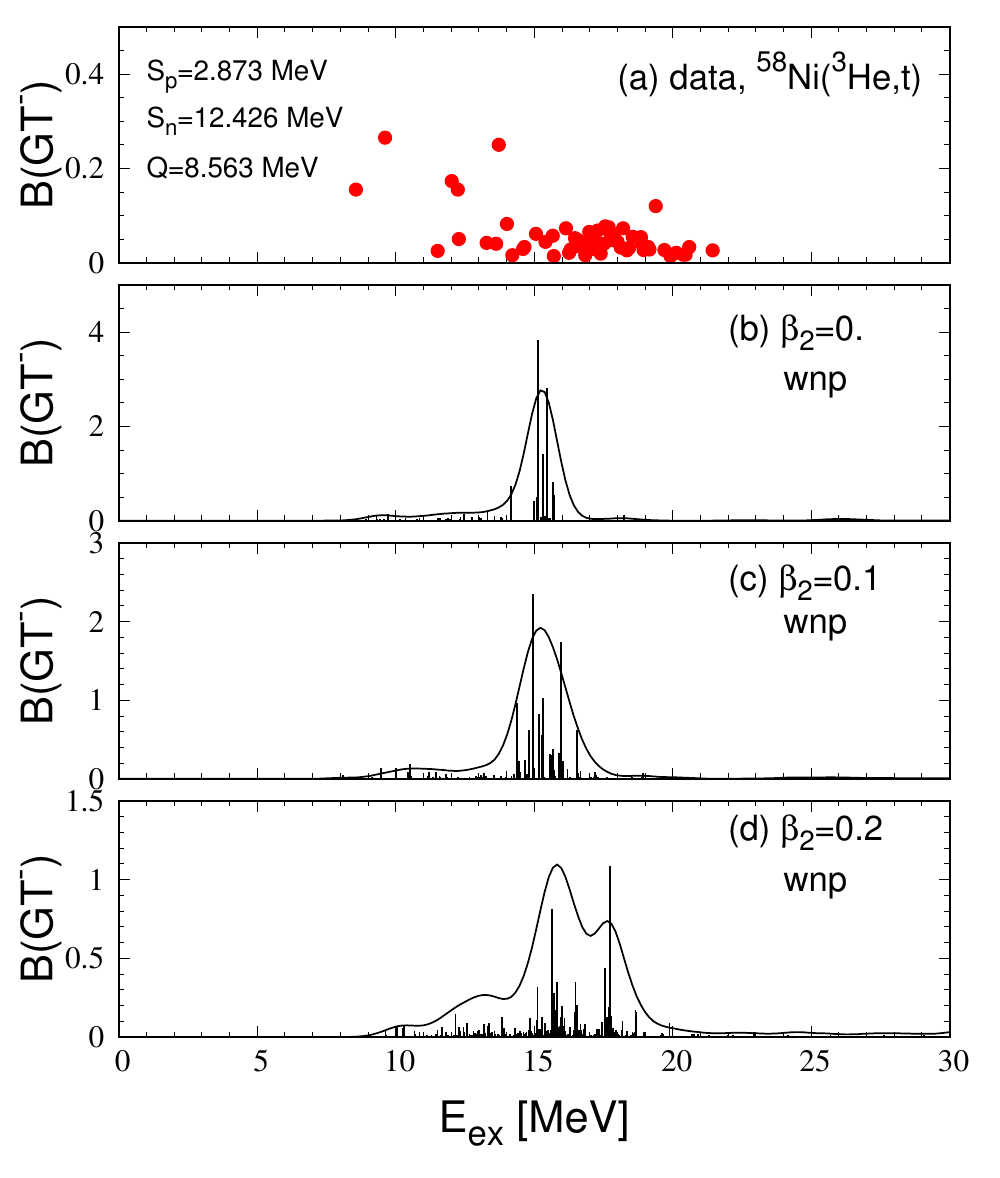}
\caption{(Color online) GT strength distributions B(GT(--)) for $^{58}$Ni.  Experimental data by $^3$He beam in panel (a) are from
Ref. \cite{Fujita07}. Left (right) panel (b)-(d) are the
results without (with) the $np$ pairing correlations,
respectively.}\label{fig5}
\end{figure}

%
In order to study the $np$ pairing effects in $N >Z$ nucleus, we
present the GT results for $^{58}$Ni in Fig. \ref{fig5}.
In Ref. \cite{Bai14}, extensive discussions of the
GT states of $pf$-shell $N = Z+2$ nuclei have been done for the
experimental data \cite{Fujita14}. They addressed
that the IS $np$ pairing could be a driving force to create low-lying GT states for those nuclei  because the IS $np$ pairing
induces only transitions with same $j$ value near to the Isobaric Analogue Resonance (IAR) owing to  the isospin operator.



If we look into detail those results in Fig.\ref{fig5}, similar effects can be found at low-lying GT states around 12 MeV region, whose distributions become spilled out to a low-lying energy region by the IS $np$ pairing contribution, but the dependence on the deformation is larger than that by the $np$ pairing and its strength by the IS $np$ pairing is small compared to the main GT peak, as explained in Ref. \cite{Bai14}.

\begin{figure}
\includegraphics[width=0.45\linewidth]{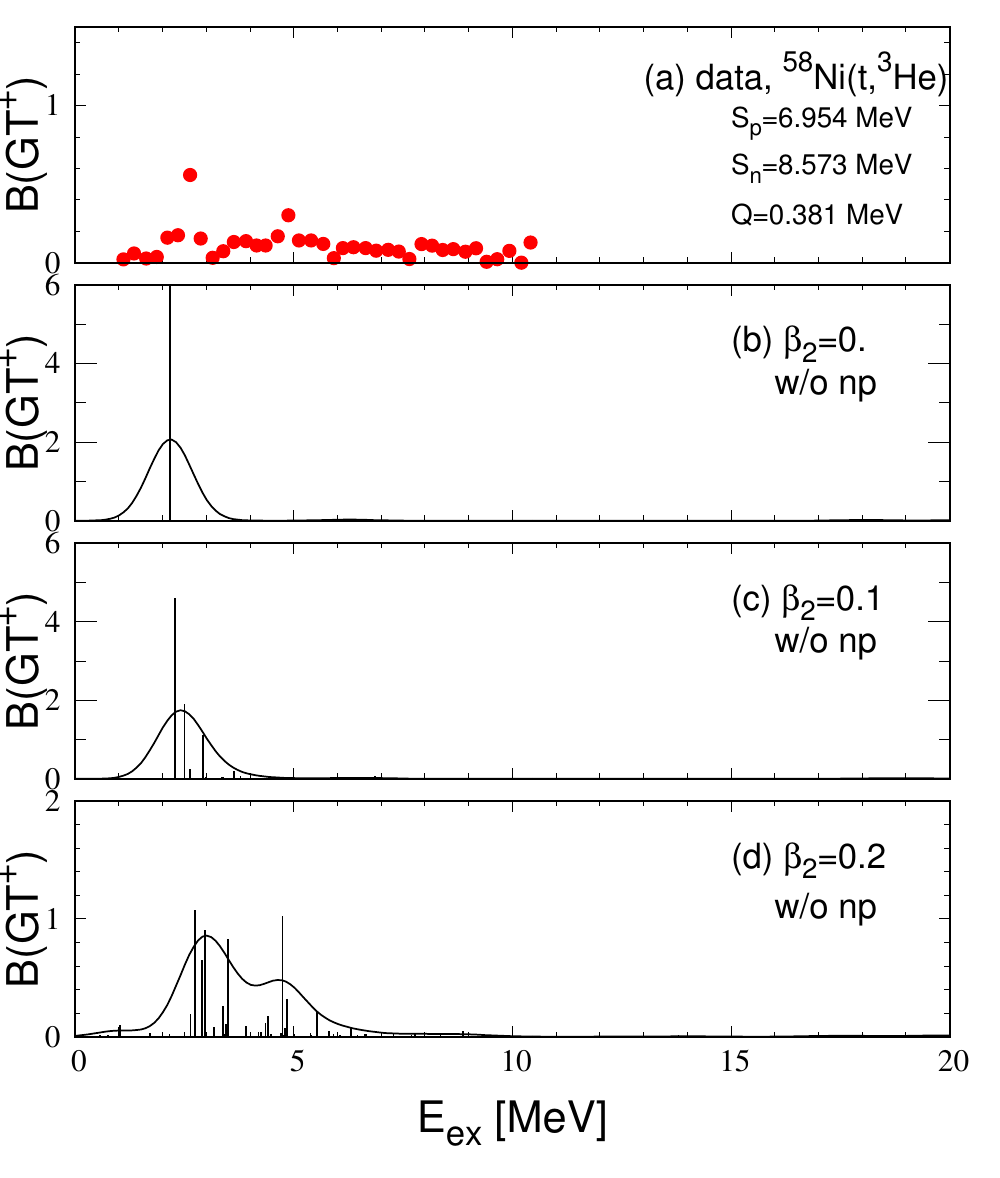}
\includegraphics[width=0.45\linewidth]{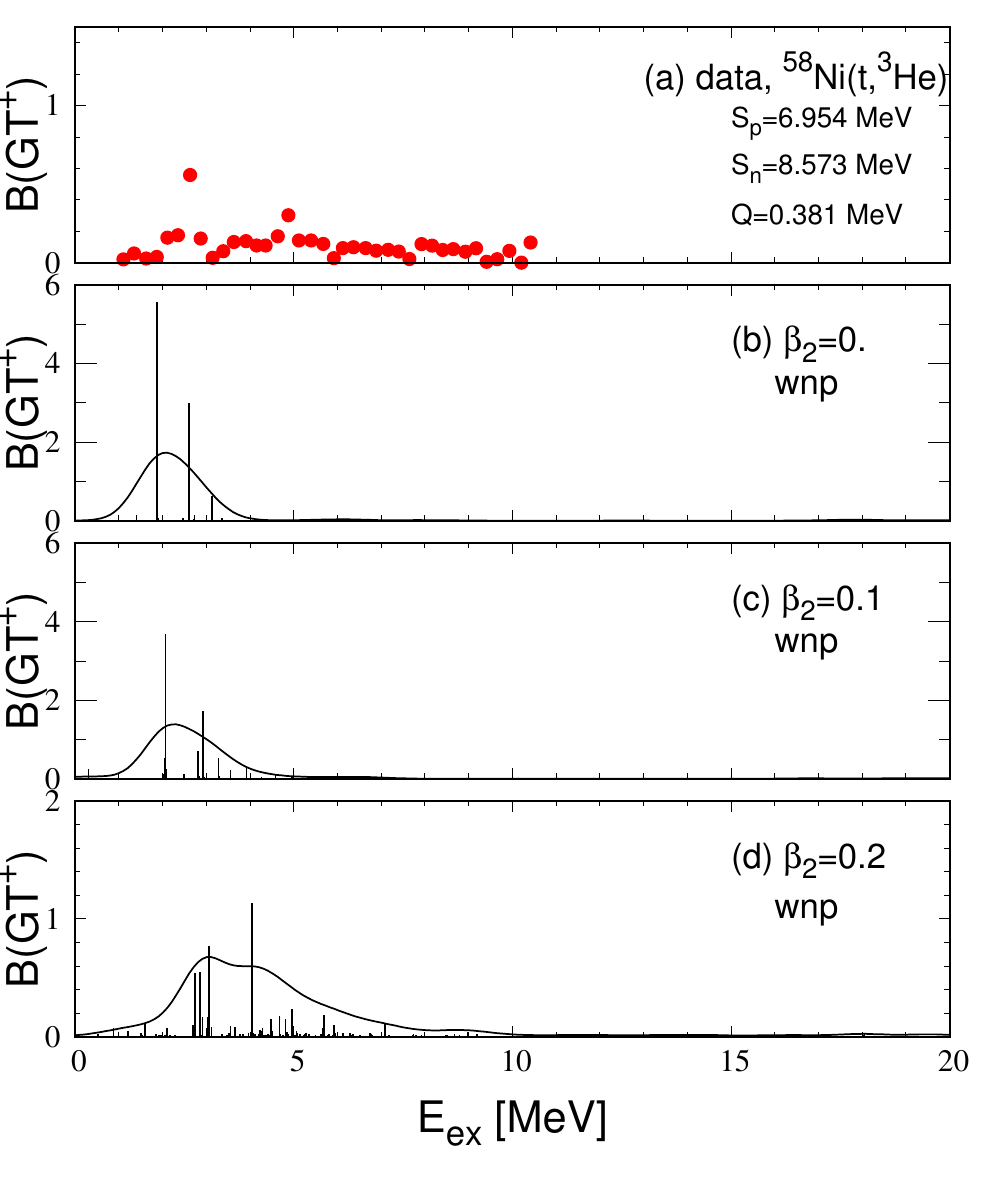}
\caption{(Color online) The same as in Fig. \ref{fig5}, but
for B(GT(+)) of $^{58}$Ni.  Experimental data by {\it t} beam in
panel (a) are from Ref. \cite{Cole06}. } \label{fig6}
\end{figure}

If we compare to the results $^{56}$Ni in Fig.\ref{fig1}, the $np$ pairing
does not show drastic effects as shown in the results
without (left) and with (right) the $np$ pairing. It means that the
$np$ pairing effects become the smaller with the increase of $N-Z$
number. However, this trend is shown to digress for $^{62,64}$Ni, as shown later on. Figure \ref{fig6} presents results of the GT(+) states and their experimental data \cite{Cole06}, where the deformation effect is more significant than the $np$ pairing effect.

\begin{figure}
\includegraphics[width=0.45\linewidth]{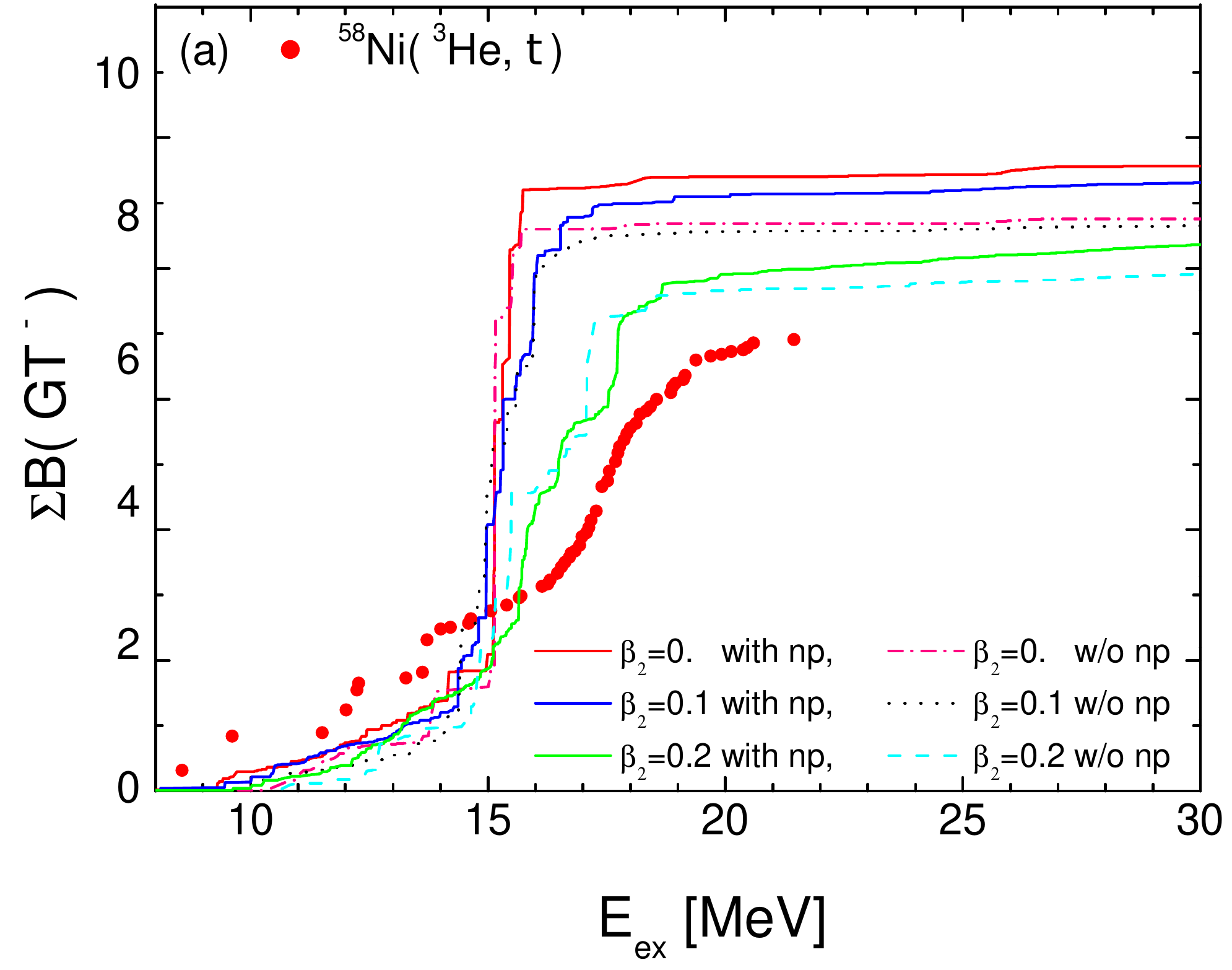}
\includegraphics[width=0.45\linewidth]{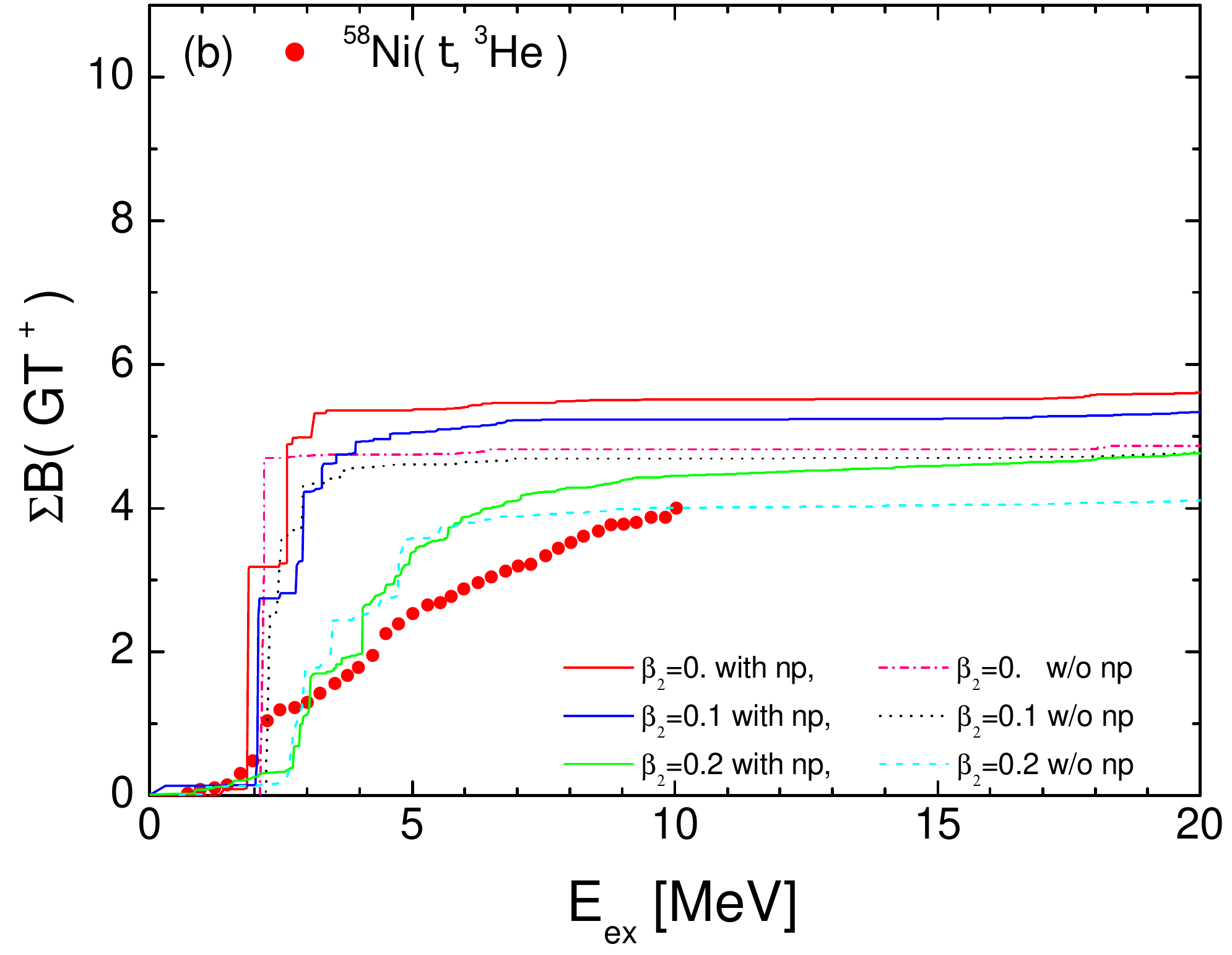}
\caption{(Color online) Running sums for the GT (--)
and GT (+) strength distributions in Figs. \ref{fig5} and \ref{fig6} (b)-(d) for $^{58}$Ni, respectively. Here
we used the general quenching factor ${(0.74)}^2$ for theoretical calculations. GT(--) data are taken from the isospin decomposition with $(p,p')$ scattering data \cite{Fujita07}. But GT(+) data are normalized to the lowest B(GT) by the calibration from $\beta$-decay \cite{Cole06}. Therefore, these data do not satisfy the Ikeda sum rule (ISR) for GT strengths. }\label{fig7}
\end{figure}
For $^{58}$Ni, the deformation effect turned out to be more
important rather than the $np$ pairing correlations, as
confirmed in the running sum in Fig. \ref{fig7}. Differences between colored solid and dot-dashed (dotted and dashed) curves by the $np$ pairing correlations are smaller than those by the deformation compared to the results in Fig. \ref{fig2} for $^{56}$Ni case. However, the $np$ pairing correlations turned out to play a role of explaining properly the GT running sum as shown in Fig. \ref{fig7}.

%
\begin{figure}
\includegraphics[width=0.45\linewidth]{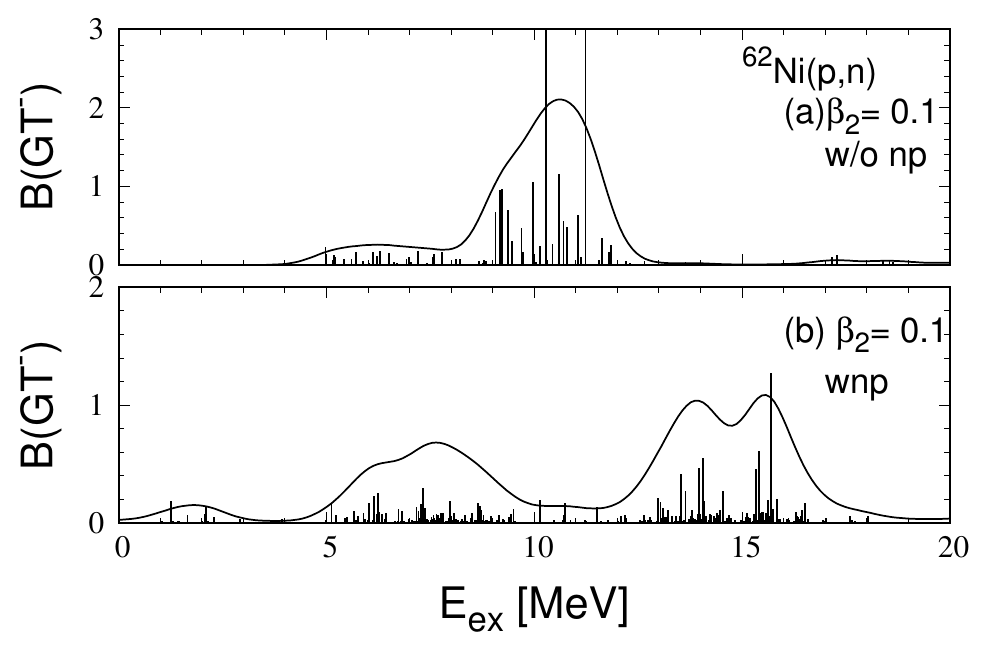}
\includegraphics[width=0.45\linewidth]{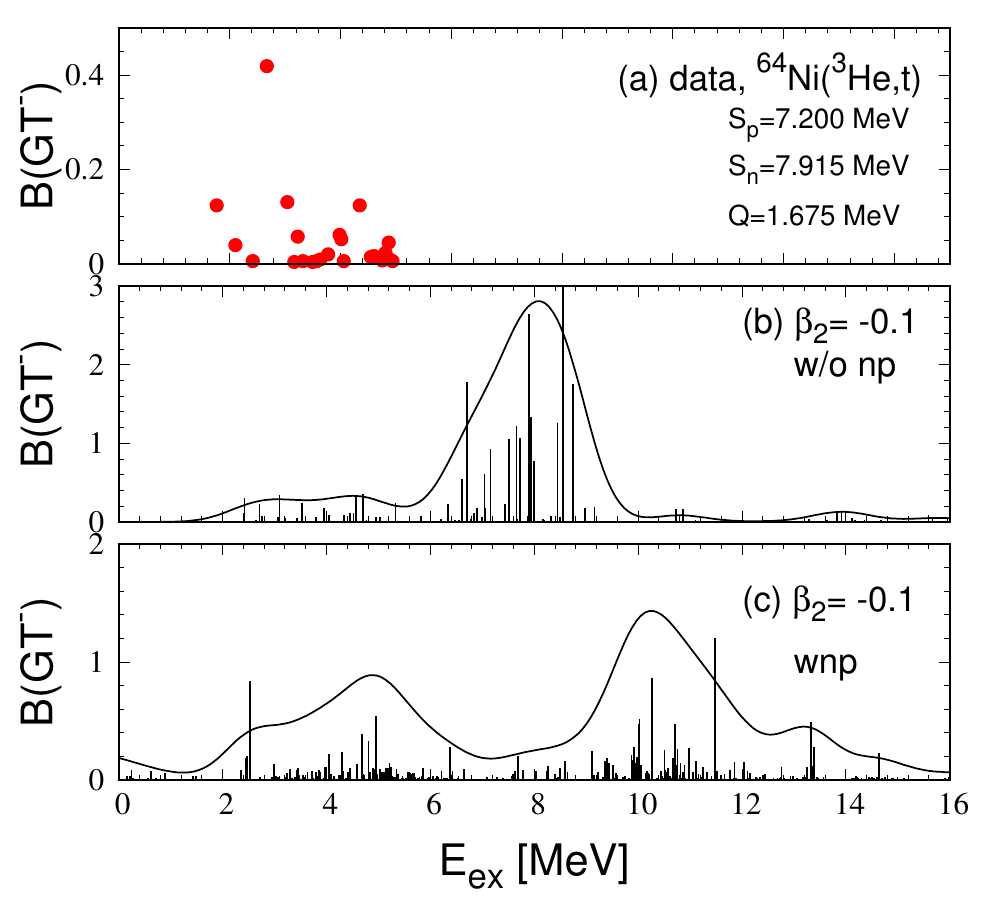}
\caption{(Color online) Same as Fig. \ref{fig5}, but for $^{62}$Ni (left) and $^{64}$Ni (right).  Experimental data for $^{64}$Ni by $^3$He beam in the right panel are from
Ref. \cite{64data}. Results are shown with and without $np$ pairing correlations, respectively.} \label{fig8}
\end{figure}

Figure \ref{fig8} provides the GT strength distributions for $^{62,64}$Ni, which shows stronger $np$ pairing effects rather than those of $^{58}$Ni in Fig.\ref{fig5}. That is, the $np$ pairing separates explicitly the GT distribution into a low-lying and a high-lying part adjacent to GTGR position. If we recollect that the $np$ pairings of $^{62,64}$Ni are almost twice of that of $^{58}$Ni as shown in Table I, this separation is quite natural.
Moreover, the larger deformation of those nuclei scatters those distributions rather than those of $^{58}$Ni. If we note the shift of the GTGR position to the reasonable region around 11 MeV, which is consistent with the data in Ref. \cite{Fujita11}, the $np$ pairing effect is indispensable even for these $pf$-shell $N \neq Z $ nuclei, $^{62,64}$Ni. More experimental data for these nuclei would be helpful for more definite conclusion for the $np$ pairing correlations effect on the GT transition strength distributions.

Recently, many papers argued the tensor force effect on the GT strength \cite{Bai14,Alex06,Urata17,Bernard16}. The tensor force effect in Ref. \cite{Bai14}, which is based on the zero range pairing force usually adopted in the Skyrme-Hartree-Fock (SHF)  approach, shifts the high-lying GT state downward a few MeV  owing to the its attractive force property. However, this trend may become changed by the deformation because of the following facts. The angular momentum $J$ is not a good quantum number in Nilsson basis, as a result it can be split by the deformation. It means that several angular momentum components are mixed in a deformed SPS, which makes direct understanding of a role of the tensor force in a deformed mean field fuzzy. Also the deformation changes the level density around the Fermi level, which leads to some increase or decrease of the pairing correlations, depending on the Fermi surface as argued in Refs. \cite{Alex06,Urata17}. For example, recent calculation in Ref. \cite{Bernard16} showed that the tensor force could be attractive, but sometimes it could be repulsive along with the SPS evolution with the deformation.

The tensor force in our DQRPA approach is explicitly taken into account on the residual interaction by the G-matrix calculated from the Bonn potential which explicitly includes the tensor force in the potential. In the mean field, the tensor force is implicitly included because of the phenomenological Woods-Saxon potential globally adjusted to some nuclear data. But, in order to discuss the separated tensor force effect on the GT strength distribution, the present approach needs more detailed analysis of the tensor force effect on the mean field as done in Ref. \cite{Alex06}, in which a tensor type potential is derived for reproducing the G-matrix calculation. But the approach is beyond the present calculation and would be a future project.

\section{Summary and conclusion}

In summary, in order to discuss the effects of the IS and IV $np$
pairing correlations and the deformation, we calculated the GT
strength distribution of $^{56,58}$Ni and $^{62,64}$Ni, which are $N=Z$ and $N-Z=$2 ,6, and 8 nucleus, respectively. The $np$ pairing effects turned out to be able
to properly explain the GT strength although the deformation was shown to be also
the other important property. In particular, the IS condensation part
played a meaningful role to explaining the GT strength distribution of $N=Z$
nucleus, $^{56}$Ni, whose GT strength distribution was shifted to a bit higher energy region by the reduction of the Fermi energy difference of proton and neutron due to the attractive $np$ pairing. For $^{58}$Ni, the deformation was more influential rather than the $np$ pairing. But for $^{62,64}$Ni, the situation is reversed because the $np$ pairing correlations are stronger than $^{58}$Ni. Namely, the $np$ pairing divides the GT strength distribution into a low and high energy region.

Therefore, the deformation treated by a mean
field approach can be balanced by the spherical property due to the
IV $np$ pairing coming from the unlike $np$ pairing as well as
the like-pairing correlations. But the IS spin-triplet $np$ part, which contributes more or less to the deformation
property due to its coupling to odd $J$ states, may give rise to
more microscopic deformation features which cannot be included in
the deformed mean field approach, and push the GT states to a bit high energy region. But all of the present results are based on a phenomenological Woods-Saxon potential. More self-consistent approaches are desirable for further definite conclusions on the IS $np$ pairing correlations. Finally, the GT strength
distribution as well as the M1 spin transition strength are shown to
be useful for investigating the IS and IV pairing properties.

\section*{Acknowledgement}
This work was supported by the National Research Foundation of Korea (Grant Nos. NRF-2015R1D1A4A01020477, NRF-2015K2A9A1A06046598, NRF-2017R1E1A1A01074023).

\end{document}